\documentclass[final]{svjour2}
\usepackage{graphicx}
\usepackage{rotating}
\usepackage{amssymb}
\usepackage{mathptmx}
\usepackage[numbers]{natbib}

\def\lesssim{\ \raise.3ex\hbox{$<$}\kern-0.8em\lower.7ex\hbox{$\sim$}\ }
\def\gesim{\ \raise.3ex\hbox{$>$}\kern-0.8em\lower.7ex\hbox{$\sim$}\ }

\makeatletter
\journalname{Journal of Low Temperature Physics}
\bibpunct{}{}{,}{s}{}{,}
\begin{document}
\newcommand{\hdblarrow}{H\makebox[0.9ex][l]{$\downdownarrows$}-}
\title{Superfluid properties of one-component Fermi gas with an anisotropic $p$-wave interaction}
\author{Daisuke Inotani$^1$ \and Manfred Sigrist$^2$ \and Yoji Ohashi$^1$}
\institute{1:Department of Physics, Keio University, Japan,
\\
2: Institut f{\"u}r Theoretische Physik, ETH Z{\"u}rich, Switzerland
\\
Tel.: +81-45-563-1141
\\ 
Fax: +81-45-566-1672
\\
\email{dinotani@rk.phys.keio.ac.jp}
}
\date{01.07.2012}
\maketitle
\begin{abstract}
We investigate superfluid properties and strong-coupling effects in a one-component Fermi gas with an anisotropic $p$-wave interaction. Within the framework of the Gaussian fluctuation theory, we determine the superfluid transition temperature $T_{\rm c}$, as well as the temperature $T_0$ at which the phase transition from the $p_x$-wave pairing state to the $p_x+ip_y$-wave state occurs below $T_{\rm c}$. We also show that while the anisotropy of the $p$-wave interaction enhances $T_{\rm c}$ in the strong-coupling regime, it suppresses $T_0$.
\end{abstract}
\section{Introduction}
Since the realization of the $s$-wave superfluid state in $^{40}$K and $^6$Li Fermi gases, the possibility of $p$-wave superfluid Fermi gas has attracted much attention both theoretically and experimentally\cite{
Regal,Ticknor,Zhang,Schunck,Gurarie,Gurarie2,Ohashi,Iskin,Botelho,Ho,Regal2,Geabler,Inaba}. A tunable $p$-wave pairing interaction associated with a $p$-wave Feshbach resonance has been realized in $^{40}$K\cite{Regal,Ticknor} and $^6$Li\cite{Zhang,Schunck} Fermi gases. It has been also observed in a $^{40}$K Fermi gas that a magnetic dipole-dipole interaction lifts the degeneracy of the $p$-wave Feshbach resonance, leading to different resonance magnetic fields between the $p_x$-component and the other $p_y$ and $p_z$ components, under an external magnetic field applied in the $x$-direction\cite{Regal,Ticknor}. This split naturally leads to the anisotropy of the three $p$-wave interaction channels as $U_x\ne U_y=U_z$ (where $U_j$ is the interaction strength in the $p_j$-channel). In this case, a phase transition from the $p_x$-wave pairing state to the $p_x+ip_y$-wave one has been theoretically predicted\cite{Gurarie,Gurarie2}. Since such a phase transition never occurs in the case of $s$-wave superfluid, the realization of the $p$-wave superfluid Fermi gas would be useful for the study of a phase transition between different pairing states, from the weak-coupling regime to the strong-coupling limit in a unified manner.
\par
Pairing fluctuations are usually suppressed in the superfluid phase, because of the opening of single-particle excitation gap. However, in the present case, even in the $p_x$-wave superfluid phase below $T_{\rm c}$, pairing fluctuations in the $p_x+ip_y$-channel would become strong near $T_0$, especially in the intermediate coupling regime. Thus, the $p$-wave superfluid Fermi gas is also an interesting system to study strong pairing fluctuations appearing in the superfluid phase.
\par
In this paper, we investigate the phase transition between the $p_x$-wave state and $p_x+ip_y$-wave state in a superfluid Fermi gas with a $p$-wave pairing interaction. So far, this problem has been examined within the Ginzburg-Landau theory\cite{Gurarie,Gurarie2}. In this paper, we employ a fully microscopic approach, including strong-coupling effects within the Gaussian fluctuation approximation\cite{Ohashi,Iskin,Botelho}. We determine the superfluid phase transition temperature $T_{\rm c}$, as well as the transition temperature $T_0$ from the $p_x$-wave state to $p_x+ip_y$-wave state below $T_{\rm c}$. 
\par

\section{Gaussian fluctuation theory for $p$-wave superfluid Fermi gas}
\par
We consider a one-component Fermi gas with a $p$-wave pairing interaction, described by the Hamiltonian
\begin{eqnarray}
H&=&\sum_{\bf p} \xi_p c_{\bf p}^{\dagger}c_{\bf p}
-\frac{1}{2}\sum_{{\bf p}{\bf p'}{\bf q}}\sum_{i=x,y,z}p_i U_i p'_i 
c_{-\bf p + \frac{\bf q}{2}}^{\dagger}c_{\bf p + \frac{\bf q}{2}}^{\dagger}
c_{\bf p' + \frac{\bf q}{2}}c_{-\bf p' + \frac{\bf q}{2}}.
\label{eq.1}
\end{eqnarray}
Here, $c_{\bf p}^\dagger$ is the creation operator of a Fermi atom with the kinetic energy $\xi_{p}=p^2/(2m)-\mu$, measured from the chemical potential $\mu$.  $-p_i U_i p'_i $ ($i=x,y,z$) are the three components of an assumed $p$-wave pairing interaction\cite{Ho}. In this paper, we ignore detailed Feshbach mechanism, and simply treat $U_i$ as a tunable parameter. However, we include the anisotropy of the interaction by the dipole-dipole interaction. That is, assuming that an external magnetic field is applied in the $x$-direction, we set $U_x>U_y=U_z$\cite{Regal,Ticknor}. 
\par
The strength of the $p$-wave interaction is conveniently measured in terms of the scattering volume $v_i$ ($i=x,y,z$) and the effective range $k_0$, that are given by, respectively,
\begin{eqnarray}
\frac{4\pi v_i}{m}&=&-\frac{U_i}{3-U_i\sum_{\bf p}^{p_{\rm c}} \frac{p^2}{2\varepsilon_p}},
\label{eq.2_1}
\\
k_0&=&-\frac{4\pi}{m^2}\sum_{\bf p}^{p_{\rm c}} \frac{p^2}{2\varepsilon_p^2}=-\frac{4}{\pi}p_{\rm c},
\label{eq.2_2}
\end{eqnarray}
where $p_{\rm c}$ is a momentum cutoff. We also introduce the anisotropy parameter, $\delta v_p^{-1}\equiv v_x^{-1}-v_y^{-1}$. 
\par
\begin{figure}
\begin{center}
\includegraphics[%
  width=0.80\linewidth,
  keepaspectratio]{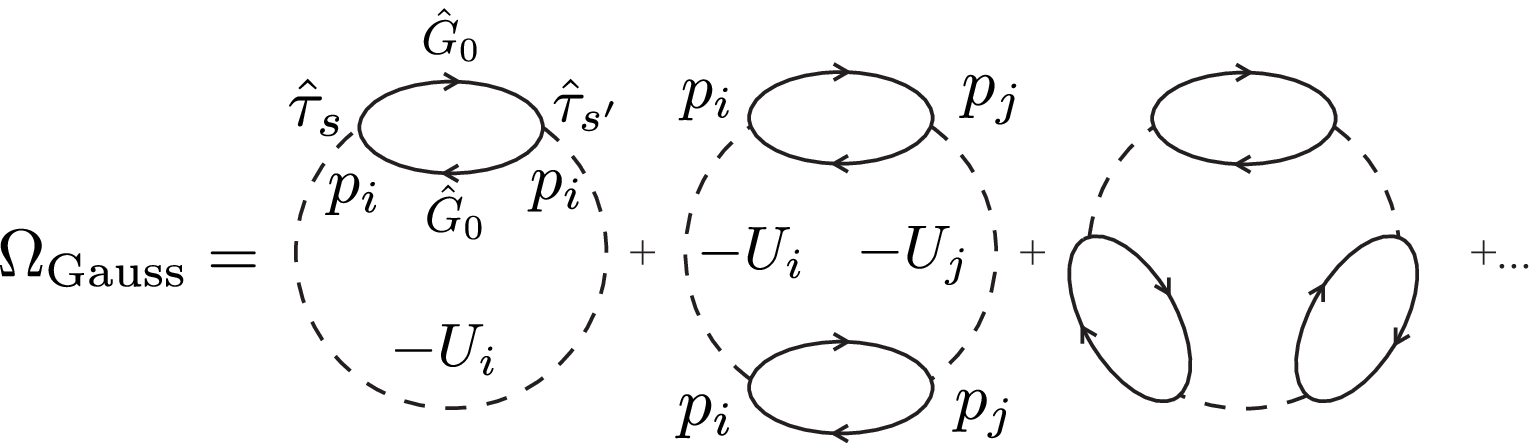}
\end{center}
\caption{Fluctuation correction $\Omega_{\rm Gauss}$ to the thermodynamic potential $\Omega$ in the $p$-wave Gaussian fluctuation theory. The solid line and the dashed line describe the $2\times2$-matrix single-particle thermal Green's function $G_0$ in the mean field theory, and the $p$-wave interaction $-p_i U_i p'_i$ ($i=x,y,z$), respectively. $\tau_{s=\pm}$ is given by $\tau_\pm=\tau_1\pm i\tau_2$, where $\tau_j$ is the Pauli matrix.}
\label{fig1}
\end{figure}
\par
We include pairing fluctuations in the $p$-wave Cooper channel within the Gaussian fluctuation theory. In this strong-coupling theory, the thermodynamic potential $\Omega$ consists of the mean field part $\Omega_{\rm MF}$ and the fluctuation part $\Omega_{\rm Gauss}$. $\Omega_{\rm MF}$ is given by
\begin{eqnarray}
\Omega_{\rm MF}&=&
\frac{1}{2} \sum_{i=x,y,z} d^*_i U^{-1}_i d_i+\frac{1}{2} \sum_{\bf p} \left[ \xi_p- E_p \right]
-\frac{1}{\beta}\sum_{\bf p} \ln\left[1+ e^{-\beta E_{p}} \right].
\end{eqnarray}
Here, ${\bf d}=(d_x,d_y,d_z)$ is the $p$-wave superfluid order parameter, and $E_{\bf p}=\sqrt{\xi_p^2 + \left| {\bf d}\cdot{\bf p} \right|^2}$ describes Bogoliubov single-particle excitations. The fluctuation part, $\Omega_{\rm Gauss}$, is diagrammatically given in Fig.\ref{fig1}. Summing up these diagrams, one has
\begin{eqnarray}
\Omega_{\rm Gauss}&=&
\frac{1}{2\beta} \ln \mathrm{det} \left[ 1+\hat{W}\hat{\pi}({\bf q},i\nu_n) \right], 
\end{eqnarray}
where $\hat{W}^{\alpha\beta}_{ij} = U_i \delta_{ij} \delta_{\alpha\beta}$ ($\alpha,\beta=1,2$ and $i,j=x,y,z$).  $\hat{\pi}^{\alpha\beta}_{ij}$ is the correlation function, having the form,
\begin{eqnarray}
\pi^{11}_{ij} \left( {\bf q},i\nu_n \right) &=&
\frac{1}{\beta}\sum_p p_i p_j\mathrm{Tr} 
\left[ \tau_{-} 
G_0 \left( {\bf p+\frac{q}{2}},i\omega_n \right)
\tau_{+} 
G_0\left( {\bf p-\frac{q}{2}}, i\omega_n-i\nu_n \right)
 \right],
\\
\pi^{12}_{ij} \left( {\bf q},i\nu_n \right) &=&
\frac{1}{\beta}\sum_p p_i p_j\mathrm{Tr} 
\left[ \tau_{-} 
G_0 \left( {\bf p+\frac{q}{2}},i\omega_n \right)
\tau_{-} 
G_0\left( {\bf p-\frac{q}{2}}, i\omega_n-i\nu_n \right)
 \right],
\\
\pi^{22}_{ij} \left( {\bf q},i\nu_n \right) &=&
{\pi^{11}_{ij}}^*\left( {\bf q},i\nu_n \right),
\\
\pi^{21}_{ij} \left( {\bf q},i\nu_n \right) &=& 
{\pi^{12}_{ij}}^*\left( {\bf q},i\nu_n \right).
\end{eqnarray}
Here, $G_0({\bf p},i\omega_n)$ is the $2\times 2$-matrix single-particle thermal Green's function in the mean field theory, given by
\begin{equation}
G_0({\bf p},i\omega_n)=
{1 \over i\omega_n-\xi_{p} \tau_3 +{\mathrm {Re}}({\bf d \cdot p}) \tau_1+ {\mathrm {Im}}({\bf d \cdot p}) \tau_2},
\label{MFG}
\end{equation}
where $\tau_j$ ($j=1,2,3$) are the Pauli matrices acting on the particle-hole space, and $\tau_{\pm} = \tau_1\pm i\tau_2$.
\par
As usual, we determine the superfluid order parameter ${\bf d}$ by solving the gap equation
\begin{eqnarray}
d_i&=& \sum_{\bf {p}} U_i p_i \frac{{\bf d \cdot p}}{2E_{\bf p}}\tanh \frac{\beta E_{\bf p}}{2},
\label{gapp}
\end{eqnarray}
together with the equation for the number $N$ of Fermi atoms,
\begin{eqnarray}
N=-\frac{\partial \Omega}{\partial \mu}
&=&\frac{1}{2}\sum_{\bf p} \left[ 1- \frac{\xi_p}{E_{\bf p}}\tanh \frac{\beta E_{\bf p}}{2} \right]
\nonumber
\\
&-&\frac{1}{2\beta} \sum_{ {\bf q},i\nu_n } \mathrm{Tr} \left[
\left(
\hat{W}^{-1}+\hat{\pi}\left( {\bf q},i\nu_n \right)
\right)^{-1} \frac{\partial \hat{\pi}\left( {\bf q},i\nu_n \right) }{\partial \mu}
\right],
\label{NUM}
\end{eqnarray}
and determine ${\bf d}$ and the Fermi chemical potential $\mu$ self-consistently.
\par
Since we are taking $U_x>U_y=U_z$, the superfluid phase transition first occurs in the $p_x$-wave Cooper channel. Thus, the equation for the superfluid phase transition temperature $T_{\rm c}$ is given by setting $i=x$ and ${\bf d}\to 0$ in Eq. (\ref{gapp}), as
\begin{equation}
1=U_x\sum_{\bf {p}} \frac{p_x^2}{2\xi_{p}}\tanh \frac{\beta \xi_{p}}{2}.
\label{TC}
\end{equation}
We solve this equation, together with the number equation (\ref{NUM}) with ${\bf q}=0$, to determine $T_{\rm c}$.

\begin{figure}
\begin{center}
\includegraphics[%
  width=0.80\linewidth,
  keepaspectratio]{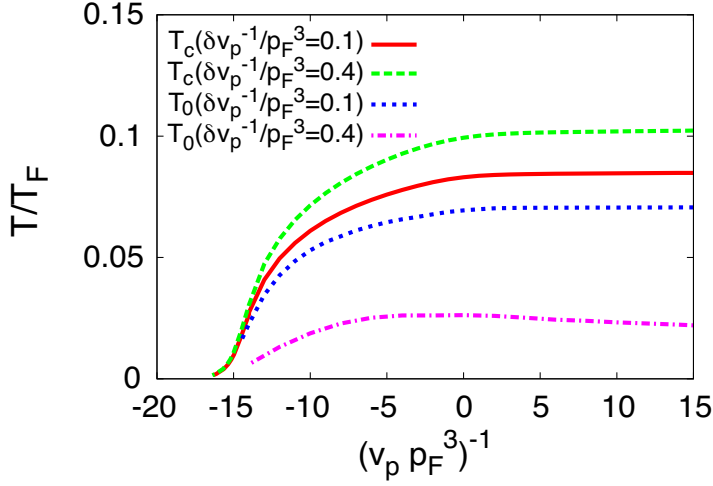}
\end{center}
\caption{(Color online) Calculated superfluid phase transition temperature $T_{\rm c}$ and the phase transition temperature $T_0$ from the $p_x$-wave state to the $p_x+ip_y$-wave state, as functions of the inverse scattering volume $(v_xp_{\rm F}^3)^{-1}$ (where $p_{\rm F}$ is the Fermi momentum). We take $k_0=-30.0p_F$.}
\label{figvp}
\end{figure}

\begin{figure}
\begin{center}
\includegraphics[%
  width=0.80\linewidth,
  keepaspectratio]{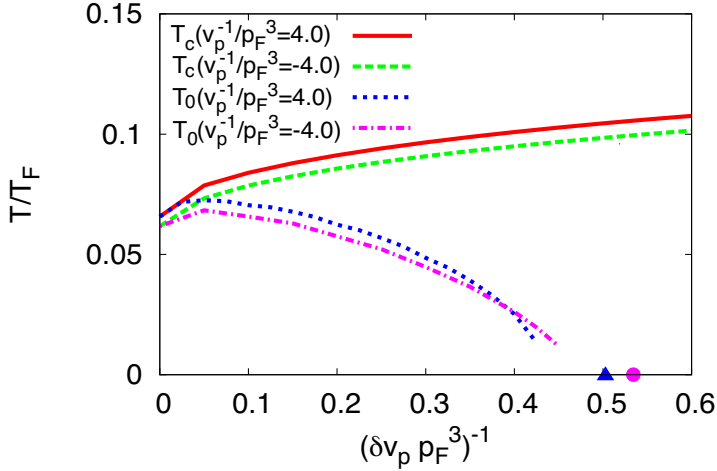}
\end{center}
\caption{(Color online) Effects of anisotropy ($\delta v_p^{-1}=v_x^{-1}-v_y^{-1}$) on the superfluid transition temperature $T_{\rm c}$ and the phase transition temperature $T_0$ from the $p_x$-wave state to the $p_x+ip_y$-wave state. The solid triangle and circle, respectively, show the critical value of $(\delta v_p p_{\rm F}^3)^{-1}$ where $T_0$ vanishes with $(v_x p_{\rm F}^3)^{-1}=4.0,-4.0$, calculated within the mean field theory.}
\label{figvpd}
\end{figure}

\par
\section{Superfluid phase transition and transition between $p_x$-wave and $p_x+ip_y$-wave states}
Figure \ref{figvp} shows $T_{\rm c}$ as a function of the interaction strength. In this figure, the increase of the inverse scattering volume $(v_xp_{\rm F}^3)^{-1}$ corresponds to the increase of the interaction strength. Starting from the weak-coupling regime, $T_{\rm c}$ gradually increases with increasing the strength of the pairing interaction, and it approaches a constant value when $(v_xp_{\rm F}^3)^{-1}\gesim 0$. Apart from the values of $T_{\rm c}$, the overall behavior of $T_{\rm c}$ is close to the $s$-wave case. 
\par
In the weak-coupling regime, Fig. \ref{figvp} shows that the anisotropy of the pairing interaction (which is described by the anisotropy parameter $\delta v_p^{-1}=v_x^{-1}-v_y^{-1}$) is not crucial for $T_{\rm c}$. In this regard, we note that, since the $T_{\rm c}$ equation (\ref{TC}) does not explicitly involve $U_y$ nor $U_z$, they only affect $T_{\rm c}$ through the Fermi chemical potential $\mu$ determined by the number equation (\ref{NUM}). However, the magnitude of $\mu$ is actually close to the Fermi energy in the weak-coupling regime because of weak pairing fluctuations. Thus, the superfluid phase transition in this regime is only dominated by $U_x$ (or $v_x$), so that $T_{\rm c}$ is insensitive to $\delta v_p^{-1}=v_x^{-1}-v_y^{-1}$.
\par
The anisotropy of the $p$-wave pairing interaction gradually becomes important, as one goes away from the weak-coupling regime. To understand this, it is convenient to consider the strong coupling limit. In this extreme case, the system may be viewed as a Bose gas, consisting of three kinds of tightly bound molecules that are formed by $U_x$, $U_y$, and $U_z$ pairing interactions. $T_{\rm c}$ is then dominated by the Bose-Einstein condensation of one of the three components having the largest number $N_{\rm B}$ of Bose molecules. While $N_{\rm B}=N/6$ in the isotropic case (where $N$ is the number of the Fermi atoms), $N_{\rm B}$ approaches $N/2$ with increasing the magnitude of $U_x$ compared with the other two interactions. Since the BEC phase transition temperature of an ideal Bose gas is proportional to $N_{\rm B}^{2/3}$, $T_{\rm c}$ increases with increasing the anisotropy parameter $\delta v_p^{-1}=v_x^{-1}-v_y^{-1}$. 
\par
Although the $p_x$-wave superfluid phase is realized near $T_{\rm c}$, this pairing symmetry changes into the $p_x+ip_y$-wave at a certain temperature ($\equiv T_0$) below $T_{\rm c}$, as shown in Fig.\ref{figvp}. While $T_{\rm c}$ is larger for a larger value of the anisotropy parameter $\delta v_p^{-1}$, $T_0$ for $(\delta v_pp_{\rm F}^3)^{-1}=0.4$ is found to be lower than that for $(\delta v_pp_{\rm F}^3)^{-1}=0.1$. To see this more clearly, we show the $(\delta v_pp_{\rm F}^3)^{-1}$-dependence of $T_0$ in Fig.\ref{figvpd}. When the $p$-wave interaction is very anisotropic ($U_x\gg U_y=U_z$), the $p_x$-wave pairing becomes more and more favorable, so that the  $p_x+ip_y$-wave state is suppressed. Although it is difficult to examine the region far below $T_{\rm c}$ based on the present strong-coupling theory because of computational problems, we briefly note that a critical value of $\delta v_p^{-1}$ at which $T_0$ vanishes can be obtained within the mean field theory.

\par
\section{Summary}
To summarize, we have investigated the superfluid properties of a one-component Fermi gas with an anisotropic $p$-wave interaction. Within the framework of the Gaussian fluctuation theory, we determined the superfluid transition temperature $T_{\rm c}$, as well as the phase transition temperature $T_0$ from the $p_x$-wave pairing state to the $p_x+ip_y$-wave state. While the anisotropy of the $p$-wave pairing interaction ($U_x>U_y=U_z$) is not crucial for $T_{\rm c}$ in the weak-coupling regime, we showed that this anisotropy enhances $T_{\rm c}$ in the strong-coupling regime. We also showed that, in contrast to the case of $T_{\rm c}$, the anisotropy of the pairing interaction suppresses $T_0$. 

\begin{acknowledgements}
We would like to thank R. Watanabe, S. Tsuchiya, S. Watabe, T. Kashimura and R. Hanai  for useful discussions. This work was supported by Grant-in-Aid from JSPS. Y. O. was supported by Grant-in-Aid for Scientific research from MEXT in Japan (22540412, 23104723, 23500056).
\end{acknowledgements}


\begin{thebibliography}{99}
\bibitem{Regal} C. A. Regal, C. Ticknor, J. L. Bohn, and D. S. Jin, Phys. Rev. Lett. {\bf 90}, 053201 (2003).
\bibitem{Ticknor} C. Ticknor, C. A. Regal, D. S. Jin, and J. L. Bohn, Phys. Rev. A {\bf 69}, 042712 (2004).
\bibitem{Zhang} J. Zhang, E. G. M. van Kempen, T. Bourdel, L. Khaykovich, J. Cubizolles, F. Chevy, M. Teichmann, L. Tarruell, S. J. J. M. F. Kokkelmans, and C. Salomon, Phys. Rev. A {\bf 70}, 030702(R) (2004).
\bibitem{Schunck} C. H. Schunck, M. W. Zwierlein, C. A. Stan, S. M. F. Raupach, W. Ketterle, A. Simoni, E. Tiesinga, C. J. Williams, and P. S. Julienne, Phys. Rev. A {\bf 71}, 045601 (2005).
\bibitem{Gurarie}V. Gurarie, L. Radzihovsky, and A. V. Andreev, Phys. Rev. Lett. {\bf 94}, 230403 (2005).
\bibitem{Gurarie2}V. Gurarie, L. Radzihovsky,  Ann. Phys. {\bf 322}, 2 (2007).
\bibitem{Ohashi} Y. Ohashi, Phys. Rev. Lett. {\bf 94}, 050403 (2005).
\bibitem{Iskin}M. Iskin and C. A. R. S\'a de Melo, Phys. Rev. Lett. {\bf 96}, 040402 (2006).
\bibitem{Botelho}S. S. Botelho and C. A. R. S\'a deMelo, J. Low Temp. Phys. {\bf 140}, 409 (2005).
\bibitem{Ho} T. L. Ho and R. B. Diener, Phys. Rev. Lett. {\bf 94}, 090402 (2005).
\bibitem{Regal2} C. A. Regal, C. Ticknor, J. L. Bohn, and D. S. Jin, Nature (London) {\bf 424}, 47 (2003).
\bibitem{Geabler} J. P. Gaebler, J. T. Stewart, J. L. Bohn, and D. S. Jin, Phys. Rev. Lett. {\bf 98}, 200403 (2007).
\bibitem{Inaba}Y. Inada, M. Horikoshi, S. Nakajima, M. Kuwata-Gonokami, M. Ueda, and T. Mukaiyama, Phys. Rev. Lett. {\bf 101}, 100401 (2008).
\end{thebibliography}
\end{document}